\documentclass{emulateapj}
\usepackage{graphics}

\begin{document}

\title{A simple parameterization of the consequences of deleptonization
for simulations of stellar core collapse}

\author{Matthias Liebend\"orfer}
\affil{CITA, University of Toronto, 60 St. George Str., Toronto,
Ontario M5S 3H8, Canada}

\begin{abstract}
A simple and computationally efficient parameterization of the deleptonization,
the entropy changes, and the neutrino stress is presented for numerical
simulations of stellar core collapse. The parameterization of the
neutrino physics is based on a bounce profile of the electron fraction
as it results from state-of-the-art collapse simulations with
multi-group Boltzmann neutrino transport in spherical symmetry. Two
additional parameters include an average neutrino escape energy and
a neutrino trapping density. The parameterized simulations
reproduce the consequences of the delicate neutrino thermalization/diffusion
process during the collapse phase and provide
a by far more realistic alternative to the adiabatic approximation,
which has often been used in the investigation of the emission of gravitational
waves, of multidimensional general relativistic effects, of the evolution
of magnetic fields, or even of the nucleosynthesis in simulations
of core collapse and bounce. For supernova codes that are specifically
designed for the postbounce phase, the parameterization builds a convenient
bridge between the point where the applicability of a stellar evolution
code ends and the point where the postbounce evolution begins.
\end{abstract}
\keywords{supernovae: general---neutrinos---radiative transfer---hydrodynamics---methods: numerical}

\section{Introduction}

Since the mid sixties, the gravitational collapse of stars has been
studied based on computer simulations \citep{Colgate.White:1966,Arnett:1966}.
Investigated topics in the collapse phase included general relativistic
dynamics \citep{May.White:1967}, magnetic fields \citep{Leblanc.Wilson:1970},
deleptonization and neutrino trapping \citep{Sato:1975,Mazurek.1976},
progenitor rotation \citep{Mueller.Hillebrandt:1981}, dissociation
energy \citep{Arnett:1982}, neutrino transport and thermalization
\citep{Wilson:1985,Bruenn:1985}, nuclear electron capture \citep{Bethe.Brown.ea:1979,Cooperstein.Wambach:1984},
and the emission of gravitational waves \citep{Moore:1981}, to name
only a few of the many possible references. Later, the field has somewhat
separated into two rather disjunct lines of research. On the one hand,
the increasing computing power was focused on the details of the neutrino
physics and neutrino transport in spherical symmetry. On the other
hand, the computational resources were invested in multidimensional
dynamics for the investigation of rotating progenitors \citep{Ott.Burrows.ea:2004,Ardeljan.Bisnovatyi.ea:2004},
general relativity \citep{Dimmelmeier.Font.ea:2002,Siebel.Font.ea:2003,Shibata.Sekiguchi:2005,Duez.Liu.ea:2005},
and magnetic fields \citep{Yamada.Sawai:2004,Kotake.Yamada.ea:2004,Liebendoerfer.Pen.Thompson:2004}.
Only few recent multidimensional collapse simulations made the effort
to include neutrino physics. These schemes are either very computationally
expensive \citep{Buras.Rampp.ea:2003,Mueller.Rampp.ea:2004,Walder.Burrows.ea:2004}
or rely on simplifications of the neutrino transport and its microphysics
\citep{Kotake.Yamada.Sato:2003,Fryer.Warren:2004}.

Among the neutrino physics that is most difficult to capture are accurate
composition-dependent rates of electron captures on nuclei \citep{Langanke.Martinez-Pinedo.ea:2003}
in the early and medium phase of collapse and the competition between
neutrino diffusion and neutrino thermalization in the later phase
of the collapse \citep{Bruenn:1985,Myra.Bludman:1989}.

Electrons at high matter density are degenerate so that they are captured
with large energies on bound or free protons.
The produced high energy neutrinos first thermalize
by electron scattering until their energy-dependent mean free path
is large enough to make diffusion competitive. As the trapped neutrinos
become degenerate themselves in the late stage of collapse, the ability
to stream away at low energy actually becomes the bottleneck for further
deleptonization. Most recent simulations of this thermalization-diffusion
process are based on an individual treatment of different neutrino
energy bins with Boltzmann neutrino transport \citep{Mezzacappa.Bruenn:1993c,Bruenn.Mezzacappa:1997,Rampp.Janka:2000,Liebendoerfer.Mezzacappa.ea:2001,Thompson.Burrows.Pinto:2003,Hix.Messer.ea:2003}.
The neutrino physics enters the equations of hydrodynamics in the
form of three different source terms: as an electron fraction change
rate, as an energy or entropy source, and as a source of acceleration
by neutrino stress.

This paper aims to bridge the two lines of research by an efficient
and very simple prescription of how published and future results of
accurate neutrino transport simulations in spherical symmetry could
be incorporated in the hydrodynamics of core collapse for a more realistic
study of the multidimensional dynamics, the role of magnetic fields,
and the emission of gravitational waves than with adiabatic simulations.

A parameterization of the three source terms is individually described
and evaluated in \( \S  \)\ref{sec:electron.fraction}-\ref{sec:velocity}.
Additional tests of the robustness of the parameterization with respect
to model changes or differences in the dynamics are conducted in \( \S  \)\ref{sec:model.dependence}.
After the conclusion, more code-specific details of the implementation
are defined in the appendix.

\section{Deleptonization}

\label{sec:electron.fraction}Stellar core collapse proceeds by an
imbalance between the self-gravitating forces of the inner core and
its fluid pressure. The baryons contribute the most significant part
to the gravitational mass of the stellar core while the degenerate
electron gas provides the dominant contribution to the pressure. The
electron fraction, defined as the number of electrons per baryon in
the gas, is therefore the most fundamental quantity for the stability
of the inner core and the evolution of its size during the dynamical
collapse. The electron fraction evolves by electron captures on nuclei
and the emission of the produced electron neutrinos. See \citep{Martinez-Pinedo.Liebendoerfer.Frekers:2004}
for a recent review of the nuclear input physics before and during
core collapse. The collapse continues until the matter at the center
reaches nuclear density. Strong interactions reduce the compressibility
at that point and the inner core bounces. The outgoing pressure wave
turns into a shock wave as soon as it reaches the sonic point at the
edge of the homologously collapsing inner core. The size of the inner core
is important because it determines the location of this transition,
the initial energy imparted to the shock, and the amount of matter
outside of the shock that will be accreted and dissociated in the
ongoing evolution.

\begin{figure}
{\centering \resizebox*{\columnwidth}{!}{\includegraphics{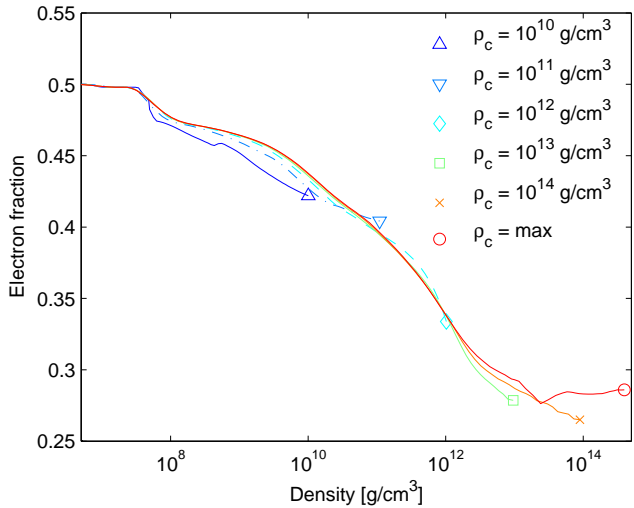} } \par}
\caption{Electron fraction profiles during core collapse in model G15 \citep{Liebendoerfer.Rampp.ea:2005}.
Each line shows the electron fraction as a function of density at
a given time. The time slices have been chosen to represent each decade
in the central density, \protect\( \rho _{c}\protect  \), as indicated
in the legend. The parameterization of the electron fraction, \protect\( Y_{e}\protect  \),
is based on the fact that the profile \protect\( Y_{e}\left( \rho \right) \protect  \)
is only a weak function of time.\label{fig:Ye.trajectories}}
\end{figure}
Figure \ref{fig:Ye.trajectories} shows the electron fraction, \( Y^{G15}_{e}\left( \rho ,t\right)  \),
as a function of density, \( \rho  \), at different times, \( t \).
The data has been taken from a general relativistic core collapse
simulation with Boltzmann neutrino transport and ``standard''
input physics.
The selected time slices correspond to the instances at which the
central density reaches \( 10^{10} \) g/cm\( ^{3} \), \( 10^{11} \)
g/cm\( ^{3} \), \( \ldots  \), \( 10^{14} \) g/cm\( ^{3} \), and
finally \( \rho _{max} \) at bounce. Figure \ref{fig:Ye.trajectories}
demonstrates that the function \( Y_{e}\left( \rho ,t\right)  \)
depends only weakly on time. Hence, it could be interesting to investigate
how hydrodynamics simulations behave when the computationally expensive
calculation of \( Y_{e}\left( \rho ,t\right)  \) is replaced by linear
interpolation in the logarithmic density of a time-independent tabulated
template of \( \bar{Y}_{e}\left( \rho \right)  \).
Because the electron fraction profile should be as accurate as possible
at the time of bounce, when the final size of the inner core is determined,
the choice \( \bar{Y}_{e}\left( \rho \right) \equiv Y^{G15}_{e}\left( \rho ,t_{b}\right)  \)
at the time of bounce, \( t_{b} \), will be investigated. The data for
the bounce electron fraction profile of model G15 are listed in
machine-readable tables in \citep{Liebendoerfer.Rampp.ea:2005}.
Alternatively, a fitting formula is provided here to increase the
flexibility of the approach. The fitting
of \( \bar{Y}_{e}\left( \rho \right)  \) is based on a piecewise
linear approximation with a piecewise cubic correction. The parameters
are two points in the \( \rho  \)-\( Y_{e} \) space, \( \left( \rho _{i},Y_{i}\right)  \),
and a scale, \( Y_{c} \), of the correction. Suggested values of the
parameters are given in Table \ref{tab:fit-formula}.
\begin{figure}
{\centering \resizebox*{\columnwidth}{!}{\includegraphics{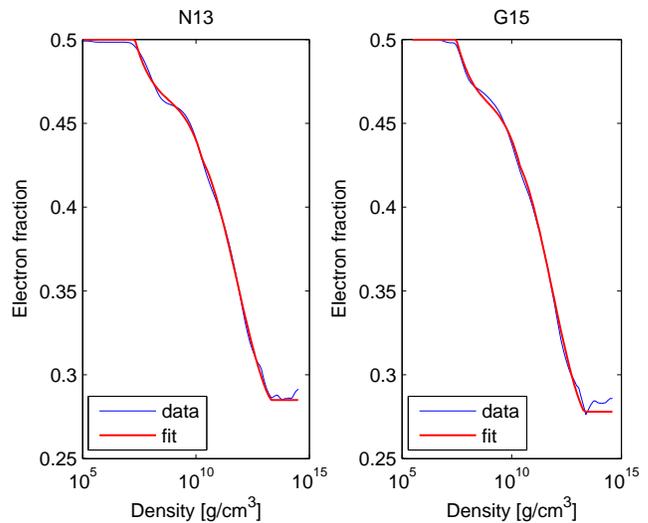} } \par}
\caption{Comparison of the fit formula for the electron fraction profile at
bounce with the original data of models N13 and G15. The agreement
is much more accurate than the deviations investigated in Fig. \ref{fig:yevol.ps}.
The rise of \protect\( Y_{e}\protect \) at the center of the G15
model is not reproduced by the fit. However, no improvement for the
simulations would be obtained, because the minimum function in Eq.
(\ref{eq:dYedt}) does not allow electron fraction increases anyway.\label{fig:fit-formula}}
\end{figure}
\begin{table}
\begin{center}
\caption{Parameters for the fitting-formula.\label{tab:fit-formula}}
\begin{tabular}{llllll}
\hline 
\hline 
& \multicolumn{2}{c}{N13} & & \multicolumn{2}{c}{G15} \\
\cline{2-3}\cline{5-6}
&
\( \rho _{i} \) {[}g/cm\( ^{3} \){]}&
\( Y_{i} \)&
&
\( \rho _{i} \) {[}g/cm\( ^{3} \){]}&
\( Y_{i} \)\\
\hline
\( i=1 \)\ldots&
\( 2\times 10^{7} \)&
\( 0.5 \)&
&
\( 3\times 10^{7} \)&
\( 0.5 \)\\
\( i=2 \)\ldots&
\( 2\times 10^{13} \)&
\( 0.285 \)&
&
\( 2\times 10^{13} \)&
\( 0.278 \)\\
\( Y_{c} \)\ldots&
&
\( 0.035 \)&
&
&
\( 0.035 \)
\\
\hline 
\end{tabular}
\end{center}
\end{table}
The fitting formula reads
\begin{eqnarray}
x\left( \rho \right)  & = & \max \left[ -1,\min \left( 1,\frac{2\log\rho -\log\rho _{2}-\log\rho _{1}}{\log\rho _{2}-\log\rho _{1}}\right) \right] \nonumber \\
Y_{e}\left( x\right)  & = & \frac{1}{2}\left( Y_{2}+Y_{1}\right) + \frac{x}{2}\left( Y_{2}-Y_{1}\right) \nonumber \\
& + & Y_{c}\left[ 1-|x|+4|x|\left( |x|-1/2\right) \left( |x|-1\right) \right] .\label{eq:fit-formula} 
\end{eqnarray}
The comparison of the fit with the original data for models N13 and
G15 in Fig. \ref{fig:fit-formula} is very satisfactory.
Table \ref{tab:fit-formula}
shows that the fits for the two models only differ in the density
at the base
of the silicon-oxygen layer, \( \rho_{1} \), and in the central electron
fraction, \( Y_{2} \).
The fit can thus easily be modified by variation of these two physical
quantities.
 
The implementation of an electron fraction evolution along \( \bar{Y}_{e}\left( \rho \right)  \)
in a pure hydrodynamics scheme is achieved by the simple prescription\begin{equation}
\label{eq:dYedt}
\frac{\delta Y_{e}}{\delta t}=\frac{\min \left( 0,\bar{Y}_{e}\left( \rho \left( t+\delta t\right) \right) -Y_{e}\left( t\right) \right) }{\delta t},
\end{equation}
where the variation is taken at the same fluid element.
The minimum function guarantees that the electron fraction monotonically
decreases even if transient instances would occur where \( \bar{Y}_{e} \)
would be larger than the actual \( Y_{e} \). At
the very start of a simulation, for example, Fig. \ref{fig:Ye.trajectories}
indicates that \( Y_{e}<\bar{Y}_{e} \) almost everywhere (\( Y_{e} \)
in the progenitor data is represented by the profile belonging to
the central density \( \rho _{c}=10^{10} \) g/cm\( ^{3} \); \( \bar{Y}_{e} \)
is represented by the profile belonging to \( \rho _{c}=max \)).
Thus, the deleptonization described by Eq. (\ref{eq:dYedt}) sets
in smoothly after a short time of adiabatic compression during which
the original electron fraction profile moves to the right in the \( \rho  \)-\( Y_{e} \)
space shown in Fig. \ref{fig:Ye.trajectories} to join the bounce
profile, \( \bar{Y}_{e} \). The latter is then followed thereafter.

\begin{figure}
{\centering \resizebox*{\columnwidth}{!}{\includegraphics{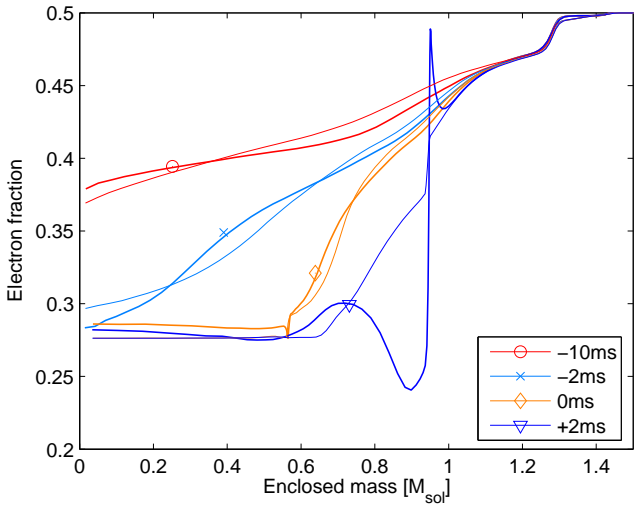} } \par}
\caption{Electron fraction profiles as functions of enclosed mass at \protect\( 10\protect  \)
ms and \protect\( 2\protect  \) ms before bounce, at bounce, and
at \protect\( 2\protect  \) ms after bounce. The thick lines represent
the accurate evolution of model G15 with Boltzmann neutrino transport.
The thin lines represent the solution with the simple parameterization
described in Eq. (\ref{eq:dYedt}). Overall, the deleptonization is
in agreement. The differences can be traced back to the earlier time
slices where deviations of up to \protect\( 5\%\protect  \) in the
deleptonization do occur. The time slice at \protect\( 2\protect  \)
ms after bounce illustrates that the simple approximation for the
deleptonization breaks down with the launch of the neutrino burst
at an enclosed mass \protect\( \sim 0.8\protect  \) M\protect\( _{\odot }\protect  \).
\label{fig:yevol.ps}}
\end{figure}

Figure \ref{fig:yevol.ps} compares the resulting evolution of the
electron fraction to the evolution obtained in the full model G15
with accurate neutrino transport. The small deviations between the
two solutions are readily explained with the profiles shown in Fig.
\ref{fig:Ye.trajectories}: At low density, early time slices (marked
by triangles) have a lower electron fraction than the bounce profile
(marked by a circle). At densities \( \sim 10^{11-12} \) g/cm\( ^{3} \)
the realistic profile (marked by a diamond) assumes a larger electron
fraction than the bounce profile. Time slices that reach to even larger
densities (marked by a square and a cross) show lower electron fractions
than the bounce profile. Because the bounce profile has been taken
as template for the parameterization, the same differences are reflected
in Fig. \ref{fig:yevol.ps}, most clearly visible in the profile at
\( 2 \) ms before bounce. The parameterized deleptonization leads
to somewhat higher \( Y_{e} \) values in the outer layers, to lower
\( Y_{e} \) values in intermediate regions, and to higher \( Y_{e} \)
values near the center. The deviations are within \( 5\% \).

The \( Y_{e} \) value in model G15 rises again around nuclear density
because of the increasing neutron degeneracy. Due to the minimum function,
this change is not captured by Eq. (\ref{eq:dYedt}) so that the central
\( Y_{e} \) value at bounce in the parameterized evolution eventually
falls below the corresponding value in the G15 model. A parameterization
of the lepton fraction instead of the electron fraction would improve
this, because the \( Y_{e} \) would then assume its correct equilibrium
value after neutrino trapping. Also the entropy changes would more
consistently derive from lepton fraction changes than from electron
fraction changes. The decision to work with the electron fraction
was motivated by the dominant role of \( Y_{e} \) in the determination
of the Chandrasekhar mass of the bouncing core, because \( Y_{e} \)
is a common input variable of realistic equations of state, and because
a numerical determination of weak equilibrium would be required to
find \( Y_{e} \) otherwise.

The \( Y_{e} \) profile of model G15 at \( 2 \) ms after bounce
shows the prominent electron fraction trough that arises when the
accretion shock dissociates matter at neutrino-transparent densities
around the enclosed mass \( \sim 0.9 \) M\( _{\odot } \) . The huge
number of neutrinos emitted in the neutrino burst also leads to neutrino
absorption ahead of the shock (manifest in the \( Y_{e} \) peak at
\( 0.95 \) M\( _{\odot } \) in the time slice at \( 2 \) ms after
bounce). These \( Y_{e} \)-changes can of course not be expected
to be represented by Eq. (\ref{eq:dYedt}) based on the stationary
bounce template. More sophisticated techniques are necessary to adequately
implement the neutrino physics in the long-term postbounce phase.

\section{Entropy changes}

\label{sec:entropy}The electron captures during collapse are not
only changing the electron fraction, the matter entropy is affected
as well. The baryons are in nuclear statistical equilibrium and the
electrons are in thermal equilibrium. Thus, the increments of the
entropy per baryon, \( \delta s \), are determined by the values
of the chemical potentials \( \mu _{n} \), \( \mu _{p} \), and \( \mu _{e} \)
of neutrons, protons, and electrons, respectively. Additionally, there
is an energy transfer between matter and neutrinos, \( \delta q \)
(e.g. \citep{Bruenn:1985} and references therein),\begin{equation}
\label{eq:entropy.general}
T\delta s=-\delta Y_{e}\left( \mu _{e}-\mu _{n}+\mu _{p}\right) +\delta q.
\end{equation}
 The temperature of the fluid is denoted by \( T \). Depending on
the density of the material and the energy of the produced neutrinos,
the neutrino can either (i) directly escape, (ii) thermalize and escape,
or (iii) be trapped for longer than the dynamical time scale.

In regime (i), \( \delta q=\delta Y_{e}\left< E\right>  \) where
\( \left< E\right>  \) is the average energy of the freely escaping
neutrinos \citep{Bethe:1990}, thus \begin{equation}
\label{eq:entropy.escape}
T\delta s=-\delta Y_{e}\left( \mu _{e}-\mu _{n}+\mu _{p}-\left< E\right> \right) .
\end{equation}
 In this regime, electron capture on nuclei dominates over electron
capture on protons. Due to the average Q-value of the nuclei \( \sim 3 \)
MeV \citep{Bruenn:1985} the neutrinos are produced with an average
energy that is only marginally larger than \( \mu _{e}-\mu _{n}+\mu _{p} \).
The corresponding small entropy decrease in this regime shall be neglected
in the following parameterization.

In regime (ii) the produced neutrinos start to be trapped by coherent
scattering off heavy nuclei. The increasing matter density causes
the electron chemical potential to rise so that the neutrinos are
produced with higher initial energies. As the neutrino mean free path
is proportional to \( E_{\nu }^{-2} \), the fastest way of escape
proceeds through thermalization at the high end of the neutrino energy
spectrum with a transition to diffusion and escape at its low end.
Figure 7 in \citep{Martinez-Pinedo.Liebendoerfer.Frekers:2004} illustrates
this thermalization process, which ends with a final escape energy
of order \( E_{\nu }^{esc}\sim 10 \) MeV. For the following parameterization
of the entropy changes, \( E_{\nu }^{esc} \) is used as a constant
parameter that defines where regime (ii) commences, namely when \( \mu _{e}-\mu _{n}+\mu _{p}-E_{\nu }^{esc}>0 \).
The entropy increase is then evaluated according to\begin{equation}
\label{eq:dsdt}
\frac{\delta s}{\delta t}=-\frac{\delta Y_{e}}{\delta t}\frac{\mu _{e}-\mu _{n}+\mu _{p}-E_{\nu }^{esc}}{T},
\end{equation}
 where \( \delta Y_{e}/\delta t \) is given by Eq. (\ref{eq:dYedt}).

At even higher densities, in regime (iii), the neutrinos are not able
to escape before bounce. They are in equilibrium with the fluid so
that \( \delta q \) in Eq. (\ref{eq:entropy.general}) is determined
by the neutrino chemical potential, \( \delta q=\delta Y_{e}\mu _{\nu } \).
The neutrinos form a normal gas component without transport abilities
and the entropy is conserved in accordance with Eq. (\ref{eq:entropy.general})
and the equilibrium condition \( \mu _{e}+\mu _{p}=\mu _{n}+\mu _{\nu }. \)
As second parameter, a density threshold \( \rho _{trap}\sim 2\times 10^{12} \)
g/cm\( ^{3} \) is introduced to define the beginning of regime (iii)
beyond which no further entropy changes are taken into account.

\begin{figure}
{\centering \resizebox*{\columnwidth}{!}{\includegraphics{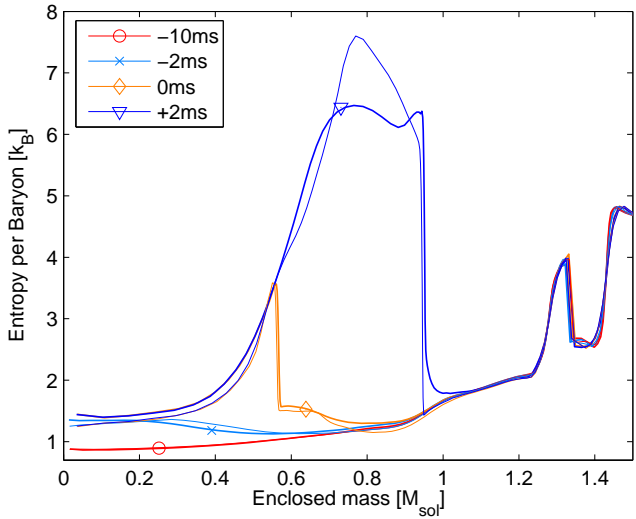} } \par}
\caption{Profiles of the entropy per baryon as functions of enclosed mass at
\protect\( 10\protect  \) ms and \protect\( 2\protect  \) ms before
bounce, at bounce, and at \protect\( 2\protect  \) ms after bounce.
The thick lines represent the accurate evolution of model G15 with
Boltzmann neutrino transport.
The thin lines represent the solution with the simple parameterization
described in Eq. (\ref{eq:dsdt}). The entropy evolution is in good
agreement during collapse, but drops to about \protect\( 7\%\protect  \)
below the reference values after bounce. The time slice at \protect\( 2\protect  \)
ms after bounce shows again the prominent difference caused by the
neutrino burst. The corresponding energy loss is not captured by the
parameterization and leads to an overestimate of the entropy around
an enclosed mass \protect\( \sim 0.8\protect  \) M\protect\( _{\odot }\protect  \).\label{fig:sevol.ps}}
\end{figure}

Figure \ref{fig:sevol.ps} presents the entropy evolution based on
Eq. (\ref{eq:dsdt}) in comparison to the entropy profiles of model
G15 with comprehensive neutrino transport. The two parameters used
in Eq. (\ref{eq:dsdt}) enable nice agreement in the collapse phase
(profiles at \( 10 \) ms and \( 2 \) ms before bounce). The density
at which the entropy starts to rise is mainly controlled by \( E_{\nu }^{esc}=10 \)
MeV. The level of the central entropy is mainly determined by the
choice of \( \rho _{trap}=2\times 10^{12} \) g/cm\( ^{3} \). The
values of the two parameters have not been particularly fine-tuned
because they should rather represent a generic estimate than a multi-digit
optimum for one particular model. The \( \sim 7\% \) lower entropy
in the parameterized evolution has no noticeable consequences for
the dynamics of the cold matter in the inner core. The entropy profiles
at \( 2 \) ms after bounce show the expected differences at late
time due to the omission of the neutrino burst in the parameterization.
The postshock region in the realistic G15 calculation is significantly
cooler because of the energy loss inferred by the neutrino emission.
The adjacent region ahead of the accretion shock is somewhat hotter
in model G15 due to neutrino absorptions.

\section{Neutrino stress}

\label{sec:velocity}Neutrino stress is the third important quantity
at the interface between neutrino transport and hydrodynamics. Although
the neutrino pressure only contributes a fraction to the gas pressure,
it influences the size of the inner core when the gas
pressure and the gravitational forces cancel to a large extent.
In regime (iii), i.e. the optically thick region where transport processes
can be neglected, the neutrino stress is determined by the gradient of
the neutrino pressure
\begin{equation}
\label{eq:acceleration.by.pressure}
\frac{d\mathbf{v}}{dt}=-\frac{\nabla p_{\nu }}{\rho }=-4\pi r^{2}\frac{\partial p_{\nu }}{\partial m},
\end{equation}
where \( d\mathbf{v}/dt \) is the Lagrangean time derivative of the
velocity. The term in the middle is the general expression; the term
on the right hand side is its spherically symmetric limit based on
an enclosed mass \( m(r) \) at radius \( r \). All
general relativistic effects shall be neglected in the derivation
of this simple parameterization. The neutrino pressure, \( p_{\nu } \),
is readily evaluated based on the thermodynamic state of the fluid,\begin{equation}
\label{eq:neutrino.pressure}
p_{\nu }=\frac{4\pi }{3\left( hc\right) ^{3}}\left( kT\right) ^{4}F_{3}\left( \frac{\mu _{\nu }}{kT}\right) .
\end{equation}
 The neutrino chemical potential is given by \( \mu _{\nu }=\mu _{e}-\mu _{n}+\mu _{p} \).
The constants \( h \), \( c \), and \( k \) refer to the Planck
constant, the speed of light, and the Boltzmann constant, respectively.
\( F_{n}\left( \eta \right) =\int _{0}^{\infty }x^{n}\left( e^{x-\eta }+1\right) ^{-1}dx \)
is the Fermi-Dirac function of order n \citep{Rhodes:1950}.

For an implementation of neutrino stress in the optically thin regime,
an estimate of the neutrino number luminosity is needed. It can be
derived from the deleptonization in Eq. (\ref{eq:dYedt})  and the
requirement of lepton conservation. The simplest
possible approximation in the construction of the neutrino flux from
distributed sources is that the neutrinos leave isotropically and
without time delay from the locations of deleptonization. Even if
this assumption is obviously wrong in core collapse,
it leads to a useful first approximation
of the non-local flux geometry \citep{Gnedin.Abel:2001}.
Firstly, the deleptonization scheme in Eq. (\ref{eq:dYedt}) already
suppresses sources at high opacities. Secondly, lepton conservation
requires that an isotropic source contributes with the square of its
inverse distance so that closeby sources from this side of
the neutrinosphere influence the direction of the neutrino flux more
strongly than remote sources from the opposite side. With this assumption,
the neutrino number flux can be expressed by the gradient of a scalar
field, \( \psi  \), which fulfills the Poisson equation\begin{equation}
\label{eq:luminosity.Poisson}
\Delta \psi =\rho N_{A}\frac{\delta Y_{e}}{\delta t}.
\end{equation}
In spherical symmetry,
the neutrino number luminosity based on Eq. (\ref{eq:luminosity.Poisson})
results in a straightforward integration of all enclosed sources,\begin{equation}
\label{eq:luminosity.estimate}
L(r)=4\pi r^{2}\frac{\partial \psi }{\partial r}=-\int _{0}^{m(r)}\frac{\delta Y_{e}}{\delta t}N_{A}dm.
\end{equation}
Avogadro's number is denoted by \( N_{A} \).

Let us stay in the spherically symmetric limit to derive an
extension of the neutrino pressure to the optically thin regions.
The neutrino stress can be expressed as a convolution of the neutrino
number luminosity, the neutrino energy, and the energy-dependent reaction
cross sections \citep{Mezzacappa.Bruenn:1993a}. If the reaction cross
sections are represented by the inverse mean free path, \( \lambda ^{-1}\left( E\right)  \),
one obtains based on the neutrino distribution function, \( f\left( E,\mu \right)  \),\begin{equation}
\label{eq:acceleration.by.stress}
\frac{dv}{dt}=\frac{2\pi }{\left( hc\right) ^{3}\rho }\int \frac{1}{\lambda \left( E\right) }f\left( E,\mu \right) E^{3}dE\mu d\mu .
\end{equation}
 The neutrino momentum phase space element, \( E^{2}dEd\mu  \), is
described by the neutrino energy, \( E \) , and the propagation angle
cosine, \( \mu  \), with respect to the radial direction. Let's now
assume that the spectrum of the number luminosity is well approximated
by a Fermi-Dirac function with degeneracy parameter \( \eta  \).
The number luminosity can then be expressed as \begin{eqnarray}
L & = & 4\pi r^{2}c\frac{2\pi }{\left( hc\right) ^{3}}\int f\left( E,\mu \right) E^{2}dE\mu d\mu \nonumber \\
 & \simeq  & 4\pi r^{2}c\frac{2\pi }{\left( hc\right) ^{3}}\left( kT_{\nu }\right) ^{3}F_{2}\left( \eta \right) H.\label{eq:luminosity} 
\end{eqnarray}
 For the following rough estimate of the decline of the neutrino stress
in the outer layers, it is further assumed that the mean free path
scales as \( \lambda ^{-1}\propto \rho E^{2} \), so that the energy
integration in Eq. (\ref{eq:acceleration.by.stress}) can be evaluated
and expressed by the number luminosity in Eq. (\ref{eq:luminosity}),\begin{equation}
\label{eq:stress.scaling}
\frac{dv}{dt}\propto \left( kT_{\nu }\right) ^{6}F_{5}\left( \eta \right) H\propto \left( kT_{\nu }\right) ^{3}\frac{F_{5}\left( \eta \right) }{F_{2}\left( \eta \right) }\frac{L}{4\pi r^{2}}.
\end{equation}
 The neutrino temperature, \( kT_{\nu } \), and the degeneracy parameter,
\( \eta  \), cannot easily be derived as a function of radius without
solving a more detailed transport problem. Though, at high opacity,
they limit to the matter temperature, \( kT \), and the neutrino
degeneracy, \( \mu _{\nu }/kT \), respectively.

The neutrino stress for the hydrodynamics simulation is now generated
by the following procedure: An estimate for the number luminosity
is given by Eq. (\ref{eq:luminosity.estimate}). Then,
at all densities larger than a specified neutrino trapping density
\( \rho _{trap} \), the neutrino stress is evaluated according to
Eqs. (\ref{eq:acceleration.by.pressure}) and (\ref{eq:neutrino.pressure}).
With the neutrino stress at density \( \rho _{trap} \), a constant
\( C \) is defined that attaches the scaling estimate found in Eq.
(\ref{eq:stress.scaling}) to the neutrino stress given by Eq. (\ref{eq:acceleration.by.pressure}),
\begin{equation}
\label{eq:stress.proportionality}
C=-4\pi r^{2}\frac{\partial p_{\nu }}{\partial m}\left[ \left( kT\right) ^{3}\frac{F_{5}\left( \mu _{\nu }/kT\right) }{F_{2}\left( \mu _{\nu }/kT\right) }\frac{L}{4\pi r^{2}}\right] ^{-1}.
\end{equation}
 In Eq. (\ref{eq:stress.proportionality}), \( r \), \( \partial p_{\nu }/\partial m \),
\( T \), \( \mu _{\nu } \), and \( L \) are evaluated at
the transition density \( \rho _{trap} \).
Now, before Eq. (\ref{eq:stress.scaling}) can be applied to extend
the neutrino stress to the region \( \rho < \rho_{trap} \), approximations
for \( kT_{\nu } \) and \( \eta  \) must be found for
regime (i). There, it is assumed that the neutrino spectrum is well
represented based on a degeneracy \( \eta =0 \). It is then consistent
with \( \S  \)\ref{sec:entropy} and the assumption of spherical symmetry
if I express the neutrino temperature
by the average neutrino energy parameter, \( E_{\nu }^{esc} \), and
set \( kT_{\nu }=F_{2}\left( 0\right) /F_{3}\left( 0\right) E_{\nu }^{esc} \).
The simplest continuous transition of the neutrino stress from regime
(iii) to regime (i) is realized by the adoption of the larger of the
two limiting cases in each intermediate point. Hence, at \( \rho < \rho_{trap} \),
the neutrino stress is approximated by \begin{equation}
\label{eq:dvdt.tail}
\frac{dv}{dt}=\frac{CL}{4\pi r^{2}}\max \left[ \left( kT\right) ^{3}\frac{F_{5}\left( \frac{\mu _{\nu }}{kT}\right) }{F_{2}\left( \frac{\mu _{\nu }}{kT}\right) },\left( E_{\nu }^{esc}\right) ^{3}\frac{F_{2}^{2}\left( 0\right) F_{5}\left( 0\right) }{F_{3}^{3}\left( 0\right) }\right] ,
\end{equation}
 where \( r \), \( T \), \( \mu _{\nu }\equiv \mu _{e}-\mu _{n}+\mu _{p} \),
and the estimate of \( L \) according to Eq. (\ref{eq:luminosity.estimate})
represent the local values at the point where \( dv/dt \) is evaluated.
Suggestions with respect to the implementation of this spherically
symmetric neutrino stress in a multidimensional hydrodynamics code
and convenient approximations for the evaluation of the Fermi integrals
are collected in the appendix.

This procedure circumvents any explicit reference to cross sections
and is therefore simple to apply. Of course, the cross sections do
enter the constant \( C \) implicitly via the relation between the
neutrino pressure gradient and the number luminosity at the transition
density \( \rho _{trap} \). As the value of the number luminosity
is derived from Eq. (\ref{eq:dYedt}), it reflects the deleptonization
rate in the reference calculation with full transport which is based
on information about the opacities used in the run that produced the
electron fraction template at bounce time.

\begin{figure}
{\centering \resizebox*{\columnwidth}{!}{\includegraphics{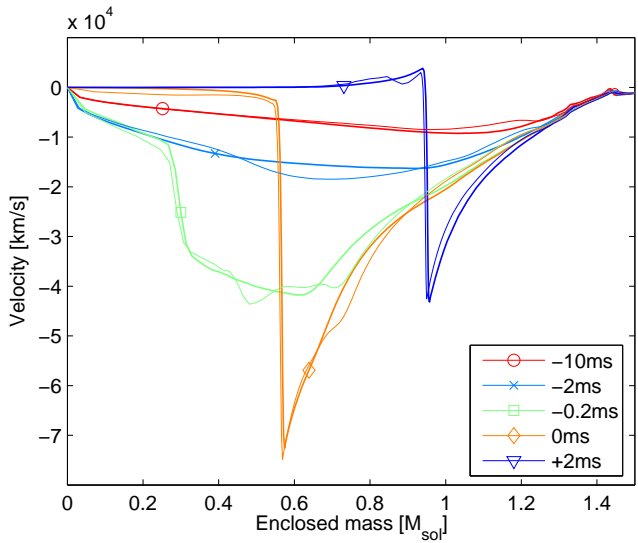} } \par}
\caption{Velocity profiles as functions of enclosed mass at \protect\( 10\protect  \)
ms, \protect\( 2\protect  \) ms, and \protect\( 0.2\protect  \)
ms before bounce, at bounce, and at \protect\( 2\protect  \) ms after
bounce. The thick lines represent the accurate evolution of model
G15 with Boltzmann neutrino transport.
The thin lines represent the solution with the simple parameterization
described in Eqs. (\ref{eq:acceleration.by.pressure}) and (\ref{eq:dvdt.tail}).
The homologous infall is in nice agreement initially. The time slice
at \protect\( 2\protect  \) ms before bounce shows a soft outgoing
pressure wave that is mainly caused by electron fraction differences
near the center. It causes perturbations around an enclosed mass of
\protect\( 0.6\protect  \) M\protect\( _{\odot }\protect  \) in
the \protect\( -0.2\protect  \) ms time slice. Otherwise, the strong
pressure wave from bounce is in good agreement with the reference
simulation at that time. The point of shock formation is very similar
in both cases. The velocity profiles even continue to coincide some
time beyond the emission of the neutrino burst as shown in the time
slice at \protect\( 2\protect  \) ms after bounce.\label{fig:vevol.ps}}
\end{figure}

Figure (\ref{fig:vevol.ps}) shows a summary of the evolution of velocity
profiles. It demonstrates that the collapse dynamics is accurately
reproduced with the parameterized neutrino physics. Some differences
can still be made out: As discussed in \( \S  \)\ref{sec:electron.fraction},
Eq. (\ref{eq:dYedt}) implies a transient halt in the deleptonization
until the density has increased enough that the initial \( Y_{e} \)
profile (labeled by \( \rho _{c}=10^{10} \) in Fig. \ref{fig:Ye.trajectories})
catches up with the template (labeled by \( \rho _{c}=max \)). This
leads to a small delay in the infall of the outer layers with respect
to the reference G15 model. This is visible in the velocity profiles
at the enclosed mass of \( 1 \) - \( 1.2 \) M\( _{\odot } \).

The time slice at \( 2 \) ms before bounce shows a soft outgoing
pressure wave in the solution with the parameterized neutrino physics
that is not present in the reference model. The first suspicion was
that it is related to the treatment of the neutrino stress because
its appearance coincides to some extent with the moment when the density
\( \rho _{trap} \) is reached at the center, i.e. the moment when
the applied neutrino stress jumps from zero to the value described
by Eq. (\ref{eq:dvdt.tail}). However, a more careful investigation
showed that the dominant reason is the shallower electron fraction
profile close to the center in the \( -2 \) ms time slice (see Fig.
\ref{fig:yevol.ps}). The outgoing wave is caused by a combination
of the excess electron pressure at the center with the electron pressure
deficit around an enclosed mass of \( 0.4 \) M\( _{\odot } \). It
introduces visible perturbations in the vicinity of the
position of shock formation
in the time slice at \( 0.2 \) ms before bounce. Nevertheless, the
much stronger outgoing pressure wave from bounce and its steepening
to the bounce-shock are nicely reproduced in the \( -0.2 \) ms and
\( 0 \) ms time slices.

The significant differences induced by the neutrino burst in the electron
fraction and entropy profiles at \( 2 \) ms after bounce seem to
have surprisingly little influence on the early dynamics. The velocity
profiles as functions of enclosed mass still agree at \( 2 \) ms
after bounce. The differences in the state of the postshock matter
rather affect the shock radius than the shock mass. While the s\(  \)hock
in the G15 model is positioned at a radius of \( 55 \) km, it already
has reached \( 63 \) km in the parameterized run due to the omission
of the lepton and energy loss by the neutrino burst.

\section{Model dependence}

\label{sec:model.dependence}After the demonstration that this simple
parameterization works well for the one model G15, this section aims
to explore to what extent the approximations are robust against variations
in the dynamics or the initial conditions. A first test is the application
of the same parameterization with the same parameters for \( E_{\nu }^{esc} \)
and \( \rho _{trap} \) to a different stellar model. The G15 model
discussed above is launched from the \citep{Woosley.Weaver:1995} model
s15s7b2. It has a quite typical iron core \( \sim 1.3 \) M\( _{\odot } \).
The run N13 in \citep{Liebendoerfer.Rampp.ea:2005} is based on the
progenitor of \citep{Nomoto.Hashimoto:1988} with an especially small
iron core \( \sim 1.17 \) M\( _{\odot }. \)
\begin{figure*}
{\centering \resizebox*{\textwidth}{!}{\includegraphics{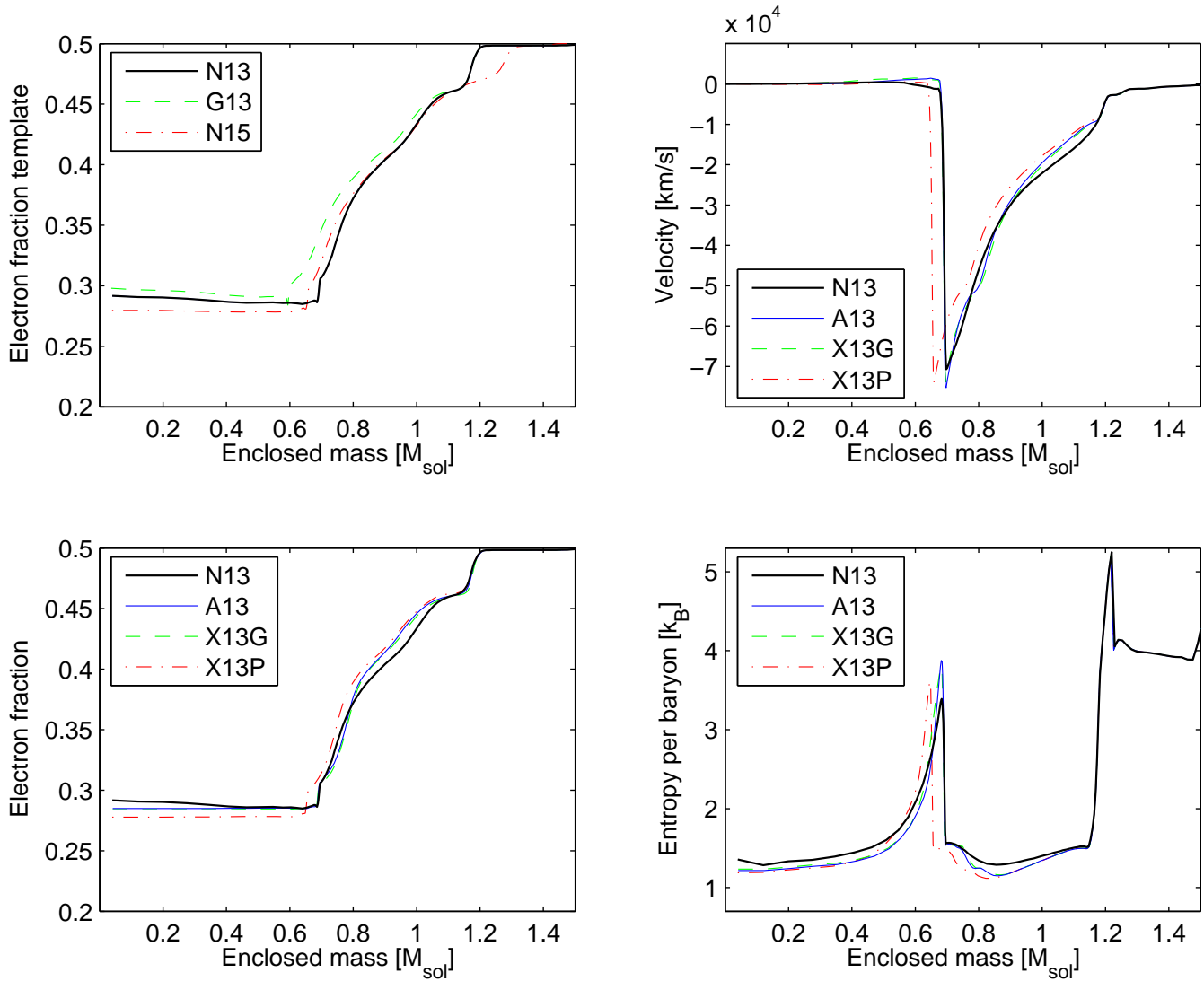} } \par}
\caption{Profiles at the time of bounce are shown to investigate the dependence
of the parameterization on model variations. (a) Electron fraction
templates for the N13, G13, and N15 model determined in simulations
with full neutrino transport. (b-d) Velocity, electron fraction, and
entropy profiles, respectively, for the parameterized simulations
at bounce. The label N13 indicates the reference results with neutrino
transport. The parameterized simulations A13, X13G, and X13P differ
only by the \protect\( Y_{e}\protect  \)-template that has been used.
Template N13 has been used for run A13, G13 has been used for run
X13G, and N15 has been used for run X13P. Swapping templates between
Newtonian and general relativistic simulations introduced no significant
differences, while a swapping of templates between progenitors lead
to slightly different positions of the shock formation that correlate
with the minimum \protect\( Y_{e}\protect  \) reached in the inner
core.\label{fig:model.ps}}
\end{figure*}

Figure \ref{fig:model.ps} compares different runs at the time of
core-bounce. Figure \ref{fig:model.ps}a shows the bounce profile
\( \bar{Y}_{e}\left( \rho \right) =Y^{N13}_{e}\left( \rho ,t_{b}\right)  \)
of the N13 run that was used to specify the deleptonization according
to Eq. (\ref{eq:dYedt}) for the parameterized run A13. The bounce
profiles of the N13 and A13 runs are displayed in Fig. \ref{fig:model.ps}b-d.
The quality of the approximation is very similar to the one discussed
above for the G15 model. The same choice of the parameters \( E_{\nu }^{esc} \)
and \( \rho _{trap} \) seems to fit also other stellar models.

The dynamics in the N13 and G15 models is also different. The former
has been calculated with Newtonian hydrodynamics and \( O(v/c) \)
neutrino transport, the latter with general relativistic dynamics
and transport. As a second test I investigate how the parameterized
solution reacts if the bounce template for a Newtonian simulation
is taken from a general relativistic model of the same progenitor
model. Thus, a bounce-template G13 in Fig. \ref{fig:model.ps}a has
been produced from a general relativistic calculation with neutrino
transport. A repetition of run A13 with the exchanged template, \( \bar{Y}_{e}\left( \rho \right) =Y^{G13}_{e}\left( \rho ,t_{b}\right)  \)
instead of \( \bar{Y}_{e}\left( \rho \right) =Y^{N13}_{e}\left( \rho ,t_{b}\right)  \),
leads to the results labeled by X13G in Fig. \ref{fig:model.ps}b-d.
In spite of the significant dynamical differences in the runs that
produced the templates N13 (solid line) and G13 (dashed line) in Fig.
\ref{fig:model.ps}a, the differences in velocity, electron fraction,
and entropy profiles between runs A13 and X13G are barely visible.
This means that the template for the parameterization of the microphysics
can be extracted from a run whose dynamics differs from the run in
which the parameterization will be applied.

A third test has been performed by the exchange of bounce templates
between the progenitor models. First, a new template has been created
with a Newtonian simulation of the s15s7b2 progenitor model labeled
by N15 in Fig. \ref{fig:model.ps}. The parameterized Newtonian run
X13P has then been launched from the \( 13 \) M\( _{\odot } \) progenitor
model using this \( \bar{Y}_{e}\left( \rho \right) =Y^{N15}_{e}\left( \rho ,t_{b}\right)  \)
template. Figure \ref{fig:model.ps}b-d shows a small deviation \( \sim 6\% \)
in the location of the shock formation. Although a given template
from one progenitor still seems to be useful for many dynamical investigations
that are launched from different progenitor models, it might be recommended
to use a template that is derived from the same progenitor to achieve
the best accuracy. A closer inspection of the differences reveals
the well-known fact that the final deleptonization in the inner core
is essential for the location of the shock formation. Because the
parameterization in Eq. (\ref{eq:dYedt}) does not support electron
fraction increases, it is the minimum \( Y_{e} \) that counts. The
minimum \( Y_{e} \) in the templates of the N13 and G13 models in
Fig. \ref{fig:model.ps}a coincide, while \( Y_{e} \) in the N15
run reached slightly lower values.

\section{Limitations}

The described parameterization has been designed to
provide a very efficient and reasonably accurate way to lead a hydrodynamics
simulation from the onset of collapse of the supernova progenitor
star to bounce. It has also been shown to be accurate for the initial
rebound of the stellar core. However, the simplicity of the described
approach entails several limitations. 

The signal of gravitational waves is not likely to be limited to the
short time interval around bounce that has traditionally been investigated
(\citet{Ott.Burrows.ea:2004} and references therein). In delayed, and especially
in neutrino-driven explosions, the signal may decay after the first
peak around bounce only to regain strength on a longer time scale
of several hundreds of milliseconds, when fluid instabilities in the
hot shocked matter between the protoneutron star and the shock front
grow to asymmetries on larger scales
\citep{Thompson:2000,Scheck.Plewa.ea:2004,Mueller.Rampp.ea:2004}.
The dynamics and the input physics for
the description of the cold matter in the collapse phase are fundamentally
different from the physics in the hydrostatic, but turbulent, hot
mantle of the protoneutron star, which eventually determines the supernova
explosion after core-bounce. The suggested scheme has not been designed
for the postbounce phase and does not describe any of the weak interaction
physics relevant after bounce.

Firstly, the scheme handles only deleptonization and not neutrino
transport. No attempt has been made to describe the very delicate
neutrino \emph{energy} deposition behind the stalled accretion shock.
Neutrino heating at \( 50 \) ms after bounce and beyond is thought
to drive important fluid instabilities. Secondly, the scheme is only applicable
to situations where the deleptonization is caused by compression of
cold matter. Even if the deleptonization in the condensing accreting
material continues to be captured after bounce, major contributions
from the emission of the neutrino burst at \( 2 \) ms after bounce
and from neutrino diffusion in the dense core are missed
by the simple parameterization. 

Therefore, if one extends an otherwise adiabatic code with the suggested
parameterization for a simulation of the \emph{postbounce} phase,
one obtains the benefit that simulations pass through a significantly
more realistic configuration at bounce and that overly fast prompt
explosions for small progenitors are avoided. But no improvement is
obtained toward an adequate description of the important neutrino
physics in the postbounce phase.

Another question is the reliability of the scheme in the collapse
phase when fast rotation rates are applied. The density distribution
is then significantly less spherically symmetric. I think one can
be somewhat optimistic because the parameterization of the deleptonization
during collapse in Eqs. (\ref{eq:dYedt}) and (\ref{eq:dsdt}) relies
more on the local density evolution than on the global geometry.
This is supported by the low sensitivity of the parameterization
to global dynamical changes investigated in \( \S \)\ref{sec:model.dependence}.
A global asymmetry changes rather the spatial distribution of local
conditions than the quality of the local conditions themselves.
Relevant local conditions are: the time scale
of compression, the density gradient, or the curvature of the isodensity
surface at the point of investigation (as a measure of the deleptonizing
volume per surface area). These quantities as function of the density may change
by several percent with respect to a spherically symmetric scenario,
but probabely not by orders of magnitude. If this speculation is true,
the parameterization would still be approximately applicable. It is
likely that accuracy is lost with increasing asphericity, but the improvement
with respect to adiabatic simulations could still be significant.

It would now be tempting to make also the treatment of the neutrino
stress multidimensional. Equation (\ref{eq:acceleration.by.pressure})
already has a multidimensional form and Eq. (\ref{eq:dvdt.tail}) is
readily generalized to
\begin{equation}
\label{eq:dvdt.tail.multiD}
\frac{d\mathbf{v}}{dt}= C\nabla \psi \max \left[ \left( kT\right) ^{3}\frac{F_{5}\left( \frac{\mu _{\nu }}{kT}\right) }{F_{2}\left( \frac{\mu _{\nu }}{kT}\right) },\left( E_{\nu }^{esc}\right) ^{3}\frac{F_{2}^{2}\left( 0\right) F_{5}\left( 0\right) }{F_{3}^{3}\left( 0\right) }\right] ,
\end{equation}
where \( \psi  \) is determined by Eq. (\ref{eq:luminosity.Poisson}).
Unfortunately, it is not clear, how \( C \) should be chosen in the
multidimensional setting. Simple choices may imply discontinuities in the
transition of the neutrino stress from Eq. (\ref{eq:acceleration.by.pressure})
to Eq. (\ref{eq:dvdt.tail.multiD}) that could introduce undesired
artefacts in the velocity field.

The spherically symmetric implementation of the neutrino stress, as layed
out in the appendix, is very simple and physically well-defined, but fails
to represent any asymmetries in the shape or temperature of the neutrinospheres
and in the emitted neutrino luminosities. This might still be an acceptable
limitation because the influence of the neutrino stress (momentum deposition)
on the collapse dynamics is not as crucial as the much discussed neutrino heating
(energy deposition) later in the postbounce phase.  
In the postbounce phase, which is not in the scope of this parameterization,
asymmetries in the neutrino field can have a significant impact on the
dynamics of the shock revival \citep{Kotake.Yamada.Sato:2003b}. On the other hand,
stellar evolution models point to rather slowly rotating inner cores
\citep{Heger.Woosley.Spruit:2005,Hirschi.Meynet.Maeder:2004} and corresponding
collapse simulations with multidimensional
neutrino transport have not shown evidence for sizable asymmetries
in the neutrino field \citep{Walder.Burrows.ea:2004}.

\section{Conclusion}

The dynamics of core collapse is dominated by electron pressure. Dynamical
simulations are only realistic if they can take account of electron
captures on bound or free protons during infall. At increasing densities
the electron captures get inhibited by neutrino phase space blocking
so that the ability to thermalize and emit the produced neutrinos
significantly contributes to the determination of the final electron
fraction in the inner core. The lower the electron fraction, the smaller
is the mass of the core that bounces when nuclear densities are reached
and the smaller is also the initial energy imparted to the outgoing
shock. These well-known relationships ask for a careful inclusion
of neutrino physics in simulations of stellar core collapse. The most
accurate treatment of neutrino transport has been developed in spherical
symmetry \citep{Mezzacappa.Messer:1999,Rampp.Janka:2002,Thompson.Burrows.Pinto:2003,Liebendoerfer.Messer.ea:2004,Sumiyoshi.Yamada.ea:2005}
where it has been coupled to the most recent evaluation of electron
capture rates on heavy nuclei \citep{Langanke.Martinez-Pinedo.ea:2003,Hix.Messer.ea:2003,Marek.Janka.ea:2005}.

The previous sections present an embarassingly simple and computationally
efficient parameterization of the deleptonization in the collapse
phase so that the consequences of past and future improvements of
the neutrino physics should be straightforward to incorporate in multidimensional
simulations that focus on other aspects of core collapse. The scheme
is based on electron fraction profiles as function of density at bounce,
\( \bar{Y}_{e}\left( \rho \right)  \), that have been produced by
spherically symmetric simulations with full neutrino transport. A fitting
formula is provided to conveniently represent these electron fraction profiles
and to allow for adjustments to future developments. It
is found that the time-evolution of the electron fraction in the simulations,
\( Y_{e}\left( \rho ,t\right)  \), follows in reasonable approximation
the profile \( \bar{Y}_{e}\left( \rho \right)  \). The corresponding
parameterization brings the electron fraction at bounce, when it is
most important for the shock dynamics, very close to the value in
the original simulation. Unfortunately, it is not exactly equal because
small differences in the time-dependence of the deleptonization will
lead to density differences at bounce so that \( \bar{Y}_{e}\left( \rho \right)  \)
is not evaluated at exactly the same position as in the original simulation.

Once the deleptonization is given, the entropy changes and neutrino
stress are estimated based on two additional parameters, an average
neutrino escape energy, \( E_{\nu }^{esc}=10 \) MeV, and a neutrino
trapping density, \( \rho _{trap}=2\times 10^{12} \) g/cm\( ^{3} \).
These parameters have not been fine-tuned, but may require adaption
if the weak interaction physics is changed.

In a comparison of the parameterized runs with the original ones it
is found that the dynamics of core collapse is quite accurately reproduced.
The parameterized neutrino physics presents a significant improvement
with respect to adiabatic treatments. Its accuracy may even rival
with neutrino transport approximations that neglect neutrino-electron
scattering.

However, some clearly visible inaccuracies do develop. The deleptonization
of central zones at high densities proceeds slower than in the reference
simulation. This leads to a weak outgoing pressure wave before bounce-density
is reached. It propagates to the point of shock formation where it
causes moderate perturbations. Inaccuracies due to a deviation in
the timing of the deleptonization are also found in the hydrodynamic
structure ahead of the shock. These deviations are a direct consequence
of the differences between the function \( \bar{Y}_{e}\left( \rho \right)  \)
and the time evolution of the electron fraction in the reference simulation.

The evolution of the electron fraction and entropy is based on the local
thermodynamic
state of the fluid. This should allow an application also in models
where the dynamics moderately differs from the spherically symmetric
case with which the electron fraction template has been produced.
The less important neutrino stress has been approximated in a spherically
averaged manner. An extension to handle asymmetries in the neutrino stress
has been sketched, but its accuracy in models with essentially aspherical
density distributions remains to be investigated.
The sensitivity of the parameterized runs to changes in the model
has been tested in two experiments that are accessible with comprehensive
neutrino transport: Firstly, a template produced with general relativistic
dynamics has been used in a parameterized run with Newtonian gravity
and, secondly, a template produced with a 15 \( M_{\odot } \) progenitor
star has been used in a parameterized run launched from a 13 \( M_{\odot } \)
progenitor star with a very different structure of the iron core.

It was found that the parameterization is not sensitive to the dynamical
differences between general relativistic and Newtonian gravity. The
choice of the progenitor for the template, however, has a small influence
on the parameterization. The extent of the deviation is directly determined
by the difference between the central \( Y_{e} \) values in the templates.
In the investigated case, the \( Y_{e} \) difference translates to
\( \sim 6\% \) difference in the point of shock formation. I thus
believe that one and the same template can be used with different
progenitors for qualitative studies of the dynamics. But a progenitor-specific
template is recommended for more quantitative investigations.

The parameterization has been designed for the collapse phase and not
for the postbounce evolution. It does not contain any physics related to
neutrino heating. During collapse, it relies on the weak interaction
physics used for the production of the bounce template. The nuclear and weak
interaction physics during collapse is an interesting field of research.
Progress in its adoption in spherically symmetric simulations has
been made \citep{Langanke.Martinez-Pinedo.ea:2003,Hix.Messer.ea:2003}
and is likely to continue. Hopefully, this paper enables the efficient
transfer of results from state-of-the-art neutrino transport simulations
in spherical symmetry to simulations with more spatial degrees of
freedom, where the implementation of comprehensive weak interaction
physics together with the accurate solution of the energy-dependent
multidimensional Boltzmann equation is yet computationally prohibitive.

\appendix
\label{sec:appendix}

\section{Implementation Details}

The simulations with the parameterized neutrino physics are based
on the same spherically symmetric general relativistic hydrodynamics
code AGILE that was used to solve the equations of hydrodynamics in
the original production of the electron fraction templates
\citep{Liebendoerfer.Rampp.ea:2005}.
The interaction between neutrino
physics and hydrodynamics proceeds through additional source terms
in the conservation equations, i.e. \( Y_{e}^{ext} \), \( e^{ext} \),
and \( S^{ext} \) in \citep{Liebendoerfer.Rosswog.Thielemann:2002}
for electron fraction change rates, total energy change rates, and
momentum change rates, respectively.
The simulations use the realistic Lattimer-Swesty equation
of state version 2.7 \citep{Lattimer.Swesty:1991}. Note that this
version is likely to crash whenever one tries to enter the nuclear
regime (eosflag=2) with a saved guess of the proton fraction obtained
from the dissociated regime (eosflag=3). As convergence in the dissociated
regime is robust, it is advisable to save guesses only when they are
returned with eosflag=2.
In the parameterized runs with AGILE, Eq. (\ref{eq:dYedt}) is used
to set \( Y_{e}^{ext} \). The heating rate \( e^{ext} \) is found by
numerical iteration of
the equation of state until the entropy change rate specified by Eq.
(\ref{eq:dsdt}) is realized. The parameter \( E_{\nu }^{esc} \)
is set to \( 10 \) MeV. Note that the approximation \( \delta s/\delta t=0 \)
in the region where \( \mu _{e}-\mu _{n}+\mu _{p}-E_{\nu }^{esc}<0 \)
will still produce a rate \( e^{ext}\neq 0 \) when the electron fraction
changes. At densities larger than \( \rho _{trap}=2\times 10^{12} \)
g/cm\( ^{3} \), the spherically symmetric limit of
Eq. (\ref{eq:acceleration.by.pressure}) is used
to calculate the neutrino stress \( S^{ext} \), while at lower densities,
Eq. (\ref{eq:dvdt.tail}) is used.
In the general relativistic simulations, the gravitational
effect of neutrino energy and pressure has only been taken into account
at densities \( \rho >\rho _{trap} \). The neutrino pressure is evaluated
by Eq. (\ref{eq:neutrino.pressure}) and the specific neutrino energy
is set to \( e_{\nu }=3p_{\nu }/\rho  \). The gravitational effect
of the neutrino luminosity was neglected.

All time derivatives, \( d/dt \), in Eqs. (\ref{eq:dYedt},\ref{eq:luminosity.estimate},\ref{eq:dsdt},\ref{eq:acceleration.by.pressure},\ref{eq:acceleration.by.stress},\ref{eq:stress.scaling},\ref{eq:dvdt.tail},\ref{eq:dvdt.tail.multiD})
are Lagrangean, i.e. taken at the same mass element. Most hydrodynamics
codes are discretized in space and not in mass; they rely on Eulerian
time derivatives, \( \partial /\partial t \). For the implementation
of above equations in these schemes I suggest to use operator splitting.
For example, the conservation equations of hydrodynamics should be
straightforward to extend with a conservation equation for electron
number,
\begin{equation}
\label{eq_electron_conservation}
\frac{\partial }{\partial t}\left( \rho Y_{e}\right) +\nabla \cdot \left( \mathbf{v}\rho Y_{e}\right) =0.
\end{equation}
In this first step, the electron fraction at location \( \mathbf{x} \)
is updated from \( Y_{e}\left( \mathbf{x},t\right)  \) to an intermediate
value \( Y_{e}^{*}\left( \mathbf{x},t+\delta t\right)  \) by the
advection of electrons. The electron fraction update is completed
in a second step by application of Eq. (\ref{eq:dYedt}),
\begin{equation}
\frac{\delta Y_{e}}{\delta t} \equiv \frac{Y_{e}\left( \mathbf{x},t+\delta t\right) -Y_{e}^{*}\left( \mathbf{x},t+\delta t\right) }{\delta t}
= \frac{\min \left( 0,\bar{Y}_{e}\left( \rho \left( \mathbf{x},t+\delta t\right) \right) -Y_{e}^{*}\left( \mathbf{x},t+\delta t\right) \right) }{\delta t}.\label{eq:dYedt.Euler}
\end{equation}
The entropy update can similarily be split into an entropy conserving
hydrodynamics update (in whatever form it is realized in the hydrodynamics
code) and a Lagrangean change of the specific entropy. For the latter,
\( \delta Y_{e}/\delta t \) in Eq. (\ref{eq:dsdt}) is substituted
by the result of Eq. (\ref{eq:dYedt.Euler}).

It is numerically stable to apply the neutrino stress in an operator
split fashion as well  because the neutrino pressure contributes
only of order \( 10\% \) to the total pressure. The simple spherical limit
in Eq. (\ref{eq:acceleration.by.pressure}) is adequate if the deviations
from spherical symmetry are small; the multidimensional form is advised
otherwise. The extension of the neutrino stress to the optically thin
regime is less straightforward
because its derivation was based on some arguments that apply only in
spherical symmetry. Eq. (\ref{eq:stress.proportionality}), for example,
extracts the opacities from the spherically
symmetric run that produced the electron fraction templates.
As long as the asphericity in the density distribution does not exceed
half a density scale height, I recommend to use spherically averaged
conditions of the multidimensional configuration to evaluate the
neutrino stress. The integration of the density over spheres results in
an enclosed mass as function of the radius. The integration of
the deleptonization in Eq. (\ref{eq:dYedt.Euler}) over spheres
leads to a luminosity estimate according to Eq. (\ref{eq:luminosity.estimate}).
Based on the spherically integrated density, energy density, and electron
density the equation of state delivers the thermodynamic conditions
used in Eq. (\ref{eq:dvdt.tail}) to derive the
neutrino stress for spheres with densities \( \rho <\rho _{trap} \).
The applicability of the multidimensional Eqs. (\ref{eq:luminosity.Poisson})
and (\ref{eq:dvdt.tail.multiD}) in highly asymmetric situations was not
numerically investigated because their assessment would require the corresponding
reference simulations with multidimensional neutrino transport.
In any case, it is important to abandon the application of the
neutrino stress in optically thin regions after bounce before
the shock reaches the density \( \rho_{trap} \) in order to
prevent the growth of the constant \( C \) in Eq.
(\ref{eq:stress.proportionality}) beyond limits.

The Fermi integrals in \( \S  \)\ref{sec:velocity} are required
for degeneracy \( \eta \geq 0 \). Convenient approximations to \( F_{2}\left( \eta \right)  \)
and \( F_{3}\left( \eta \right)  \) have been taken from \citep{Epstein.Pethick:1981},
while \( F_{5}\left( \eta \right)  \) has been derived along similar
lines based on \citep{Rhodes:1950}: \begin{eqnarray}
F_{2}\left( \eta \right)  & \simeq  & \frac{1}{3}\left( \eta ^{3}+\pi ^{2}\eta \right) +\frac{3}{2}\zeta \left( 3\right) e^{-\alpha \eta }\nonumber \\
F_{3}\left( \eta \right)  & \simeq  & \frac{1}{4}\left( \eta ^{4}+2\pi ^{2}\eta ^{2}+\frac{7\pi ^{4}}{15}\right) -\frac{7\pi ^{4}}{120}e^{-\eta }\nonumber \\
F_{5}\left( \eta \right)  & \simeq  & \frac{1}{6}\left( \eta ^{6}+5\pi ^{2}\eta ^{4}+7\pi ^{4}\eta ^{2}+\frac{31\pi ^{6}}{21}\right)
 - \frac{31\pi ^{6}}{252}e^{-\eta }\label{eq:fermi.integrals} 
\end{eqnarray}
 with \( \zeta \left( 3\right) \simeq 1.202 \) and \( \alpha =2\pi ^{2}/\left( 9\zeta \left( 3\right) \right) -1\simeq 0.825 \).


\end{document}